\documentclass{llncs}

\usepackage{multirow}

\usepackage{subfig}
\usepackage{caption}
\usepackage{subfloat}

\usepackage[pdftex]{graphicx}
\usepackage{epstopdf}

\usepackage{soul} % text highlighting
\usepackage{xcolor} % to highlight text

\begin{document}
\bibliographystyle{plain}

\title{Weighted Random Sampling over Data Streams}

\author{Pavlos S. Efraimidis}

\institute{Department of Electrical and Computer Engineering\\
Democritus University of Thrace, Xanthi, Greece \\
pefraimi@ee.duth.gr}

\maketitle

\begin{abstract}
In this work, we present a comprehensive treatment of weighted random sampling (WRS)
over data streams. More precisely, we examine two natural interpretations
of the item weights, describe an existing algorithm for each case (\cite{Ch82,ES06}),
discuss sampling with and without replacement and show adaptations of the algorithms
for several WRS problems and evolving data streams.
\end{abstract}

\section{Introduction}
The problem of random sampling calls for the selection of $m$
random items out of a population of size $n$. If all items have
the same probability to be selected, the problem is known as uniform random sampling.
If each item has an associated weight and the probability of each item to be selected is 
determined by these item weights, then the problem is called weighted random sampling (WRS).

Weighted random sampling, and random sampling in general, is a fundamental problem with applications
in several fields of computer science including databases, data streams, data mining
and randomized algorithms. Moreover, random sampling is important
in many practical problems, like market surveys, quality control in
manufacturing, statistics and on-line advertising.

There are several factors that have to be taken into account, when facing a WRS problem. 
It has to be defined if the sampling procedure is with or
without replacement, whether the sampling procedure has to be executed
over data streams, and what the semantics of the item weights are. 
In this work, we present a comprehensive treatment of WRS
over data streams. In particular, we examine the above problem parameters
and describe efficient solutions for different WRS problems that
arise in each case.
\begin{itemize}
\renewcommand{\labelitemi}{$\circ$}
\item {\bf Weights.}
In WRS the probability of each item to
be selected is determined by its weight with respect to the weights of
the other items. However, for random sampling schemes
without replacement there are at least two natural ways to
interpret the item weights.
In the first case, the relative weight of each item determines the
probability that the item is in the final sample. In the second,
the weight of each item determines the probability that the item
is selected in each of the explicit or implicit item selections
of the sampling procedure. Both cases will become clear in the sequel.

\item {\bf Replacement.} Like other sampling procedures, the WRS procedures
can be with replacement or without replacement.
In WRS with replacement, each selected item is replaced in the main lot with
an identical item, whereas in WRS without replacement
each selected item is simply removed from the population.
\item {\bf Data Streams.} Random sampling is often applied to very large datasets and in particular
to data streams. In this case, the random sample has to be generated
in one pass over an initially unknown population.
An elegant and efficient approach to generate random samples from
data streams is the use of a reservoir of size $m$, where $m$ is the sample
size. The reservoir-based sampling algorithms maintain the invariant that,
at each step of the sampling process, the contents of the reservoir
are a valid random sample for the set of items that have been processed up to
that point. There are many random sampling algorithms that make use
of a reservoir to generate uniform random samples over data streams~\cite{Vi85}.
\item {\bf Feasibility of WRS.}
When considering the problem of generating a weighted random sample
in one pass over an unknown population one may doubt that this is possible.
In~\cite{Ag06}, the question whether reservoir maintenance can
be achieved in one pass with arbitrary bias functions, is stated as an open problem.
In this work, we bring to the fore two algorithms~\cite{Ch82,ES06}
for the two, probably most important, flavors of the problem.
In our view, the above results, and especially the older one,
should become more known to the databases and algorithms communities.
\item {\bf A Standard Class Implementation.} Finally, we believe that the algorithms
for WRS over data streams can and should be part of standard class
libraries at the disposal of the contemporary algorithm or software engineer.
To this end, we design an abstract class for WRS and provide prototype
implementations of the presented algorithms in Java.
\end{itemize}

\noindent
{\bf Contribution and Related Work.}
Random sampling is a classic, well studied field, and the volume of the corresponding literature is enormous.
See for example~\cite{Kn81,Vi84,Vi85,Ol93,Li94} and the references therein. 
These results concern uniform random sampling, random sampling with a reservoir (which can be used
on data streams), and weighted random sampling but not over data streams. 
An efficient algorithm for weighted random sampling with a reservoir which can support data streams 
is presented in~\cite{ES06}.
Another weighted random sampling algorithm, which is less known to the computer science community
and which uses a different interpretation for the item weights, is presented in~\cite{Ch82}. 
This algorithm, too, is efficient and can be applied to data streams. 

Random sampling is still an active research field and new sampling schemes are studied in
various contexts; some indicative examples are sampling from sliding windows~\cite{Lo07},
from distributed data streams~\cite{Co10,Ti11,Co12}, from streams with time decay~\cite{Co09},
independent range sampling~\cite{Hu14}, sampling on very large file systems~\cite{Go14},
and stratified reservoir sampling~\cite{AlKateb2014}. In light of the above results (which are mainly from the data streams field), 
we consider the algorithms of~\cite{Ch82} and~\cite{ES06} as fundamental sampling schemes
for general purpose weighted random sampling over data streams. 

In this work, we present a comprehensive treatment of general purpose weighted random sampling (WRS) 
over data streams. More precisely, we identify and examine two natural interpretations
of the item weights, describe an existing algorithm for each case (\cite{Ch82,ES06}),
discuss sampling with and without replacement and show adaptations of the algorithms
for several WRS problems and evolving data streams. Moreover, we bring to the fore the 
sampling algorithm of Chao and show how to apply the jumps technique on it.
Finally, we propose an abstract class definition for weighted random sampling over data streams
and present a prototype implementation for this class.

\noindent
{\bf Outline.} The rest of this work is organized as follows:
Notation and definitions for WRS problems
are presented in Section~\ref{sec:WRS}.
Core algorithms for WRS are described in~\ref{sec:corealg}.
The treatment of representative WRS problems is described in~\ref{sec:alg}.
In Section~\ref{sec:impl}, a prototype implementation and experimental results
are presented. Finally, the role of item weights is examined in~\ref{sec:weights}
and an overall conclusion of this work is given in~\ref{sec:disc}.

\section{Weighted Random Sampling (WRS)}
\label{sec:WRS}

Given an instance of a WRS problem, let $V$ denote the population
of all items and $n = |V|$ the size of the population.
In general, the size $n$ will not be known to the WRS algorithms.
Each item $v_i \in V$, for $i=1,2,\dots,n$, of the population has an associated
weight $w_i$. The weight $w_i$ is a strictly positive real number $w_i > 0$
and the weights of all items are initially considered unknown.
The WRS algorithms will generate a weighted random sample of size $m$.
If the sampling procedure is without replacement then it must hold that $m \leq n$.
All items of the population are assumed to be discrete, in the sense
that they are distinguishable but not necessarily different.
The distinguishability can be trivially achieved by assigning an increasing ID
number to each item in the population, including the replaced items (for WRS with replacement).
We define the following notation to represent the various WRS problems:
\begin{equation}
WRS-\mathrm{<rep>}-\mathrm{<role>},
\end{equation}
where the first parameter specifies the replacement policy
and the second parameter the role of the item weights.

\begin{itemize}
\renewcommand{\labelitemi}{$\bullet$}
\item Parameter {\bf rep:} This parameter determines if and how many times a selected item
can be replaced in the population. A value of ``N'' means that each selected item is not replaced
and thus it can appear in the final sample at most once, i.e., sampling without replacement.
A value of ``R'' means that the sampling procedure is with replacement
and, finally, an arithmetic value $k$, where $1 \leq k \leq m$, defines that each item
is replaced at most $k-1$ times, i.e., it can appear in the final sample at most $k$ times.
\item Parameter {\bf role:}  This parameter defines the role of the item weights
in the sampling scheme. As already noted, we consider two natural ways to interpret
item weights. In the first case, when the role has value P, the probability of an item
to be in the random sample is proportional to its
relative weight. In the second case, the role is equal to W and the relative weight determines
the probability of each item selection, if the items would be selected sequentially.
\end{itemize}

Moreover, WRS-P will denote the whole class of WRS problems where the item weights
directly determine the selection probabilities of each item, and WRS-W
the class of WRS problems where the items weights determine the selection
probability of each item in a supposed\footnote{
We say ``supposed'' because even though WRS is best described
with a sequential sampling procedure, it is not inherently sequential.
Algorithm A-ES~\cite{ES06} which we will use to solve WRS-W problems
can be executed on sequential, parallel and distributed settings.}
sequential sampling procedure. A summary of the notation
for different WRS problems is given in Table~\ref{tab:notation}.

\begin{table}[ht]
\begin{minipage}[b]{\linewidth}
\centering
{\footnotesize
\begin{center}
\begin{math}
\begin{array}{|c|c|c|}
\hline \multicolumn{2}{|c|}{\mbox{\bf WRS Problem}} & \mbox{\bf Notation} \\
\hline \multicolumn{2}{|c|}{\mbox{ With Replacement}} & \mbox{ WRS-R } \\
\hline \multirow{2}{*}{\mbox{ Without Replacement }} & \mbox{ Probabilities } & \mbox{ WRS-N-P } \\ \cline{2-3}
 & \mbox{ Weights } & \mbox{ WRS-N-W } \\
\hline \mbox{ With $k-1$ Replacements } & \mbox{ Weights } & \mbox{ WRS-k-W } \\
\hline
\end{array}
\end{math} \\
\end{center}
} \caption[Notation]{Notation for WRS problems.}
\label{tab:notation}
\end{minipage}
\end{table}

\begin{definition}
\label{def:WRS-m}
Problem WRS-R (Weighted Random Sampling with Replacement). \\
Input: A population of $n$ weighted items and a size $m$ for the random sample.\\
Output: A weighted random sample of size $m$. The probability of each item to occupy
each slot in the random sample is proportional to the relative weight of the item, i.e.,
the weight of the item with respect to the total weight of all items.
\end{definition}

\begin{definition}
\label{def:WRS-P-1}
Problem WRS-N-P (Weighted Random Sampling without Replacement, with defined Probabilities). \\
Input: A population of $n$ weighted items and a size $m$ for the random sample.\\
Output: A weighted random sample of size $m$. The probability of each item to be included
in the random sample is proportional to its relative weight.
\end{definition}

\noindent
Intuitively, the basic principle of WRS-N-P can be shown with the following example.
Assume any two items $v_i$ and $v_j$ of the population with weights $w_i$ and $w_j$, respectively.
Let $c = w_i/w_j$. Then the probability $p_i$ that $v_i$ is in the random sample
is equal to $c \, p_j$, where $p_j$ is the probability that $v_j$ is in the random sample.
For heavy items with relative weight larger than $1/m$ we say that the respective items
are ``infeasible'' or ``overweight''. If the inclusion probability of an overweight item would be
proportional to its weight, then this probability would become larger than 1, which of course
is not possible. As shown in Section~\ref{sec:A-Chao}, the overweight items are handled
in a special way that guarantees that they are selected with probability exactly $1$.

\begin{definition}
\label{def:WRS-W-1}
Problem WRS-N-W (Weighted Random Sampling without Replacement, with defined Weights). \\
Input: A population of $n$ weighted items and a size $m$ for the random sample.\\
Output: A weighted random sample of size $m$. In each round, the probability of every
unselected item to be selected in that round is proportional to the relative item weight
with respect to the weights of all unselected items.
\end{definition}

\noindent
The definition of problem WRS-N-W is essentially the following sampling procedure.
Let $S$ be the current random sample. Initially, $S$ is empty.
The $m$ items of the random sample are selected in $m$ rounds.
In each round, the probability for each item in $V-S$ to be selected is
$p_i(k) = \frac{w_i}{\sum_{s_j \in V-S} w_j}$. Using the probabilities
$p_i(k)$, an item $v_k$ is randomly selected from $V-S$ and inserted into $S$.
We use two simple examples to illustrate the above defined WRS problems.

\begin{example}
Assume that we want to select a weighted random sample
of size $m=2$ from a population of $n=4$ items with weights $1, 1, 1$ and $2$, respectively.
For problem WRS-N-P the probability of items 1, 2 and 3 to be in the random sample
is $0.4$, whereas the probability of item 4 is $0.8$.
For WRS-N-W the probability of items 1, 2 and 3 to be in the random sample
is $0.433$, while the probability of item 4 is $0.7$.
\end{example}

\begin{example}
\label{exa:infeasible}
Assume now that we want to select $m=2$ items from a population of $4$ items
with weights $1, 1, 1,$ and $4$, respectively.
For WRS-N-W the probability of items $1$, $2$ and $3$ to be in the random sample
is $0.381$, while the probability of item $4$ is $0.857$.
For WRS-N-P, however, the weights are infeasible because the weight of item $4$
is infeasible. In particular, the product $m$ times the relative weight of item $4$ is 
$2 \cdot (4/7)$ which is larger than $1$ and cannot be used as a probability.
This case is handled by assigning with probability 1 a position
of the reservoir to item 4 and filling the other position of the reservoir randomly
with one of the remaining (feasible) items. Note that if the sampling procedure
is applied on a data stream and a fifth item, for example with weight 3, arrives, then
the instance becomes feasible with probabilities $0.2$ for items 1, 2 and 3,
$0.8$ for item 4 and $0.6$ for item $5$. The possibility for infeasible
problem instances or temporary infeasible evolving problem instances over
data streams is an inherent complication of the WRS-N-P problem that has to
be handled in the respective sampling algorithms.
\end{example}

\section{The Two Core Algorithms}
\label{sec:corealg}

The two core algorithms that we use for the WRS problems
of this work are the \emph{General Purpose Unequal Probability Sampling Plan}
of Chao~\cite{Ch82} and the \emph{Weighted Random Sampling with a Reservoir}
algorithm of Efraimidis and Spirakis~\cite{ES06}. We provide a short description
of each algorithm while more details can be found in the respective papers.

\subsection{A-Chao}
\label{sec:A-Chao}
The sampling plan of Chao~\cite{Ch82}, which we will call A-Chao, is a reservoir-based
sampling algorithm that processes sequentially an initially unknown
population V of weighted items.

A typical step of algorithm A-Chao is presented in Figure~\ref{fig:A-Chao}.
When a new item is examined, its relative weight is calculated and used
to randomly decide if the item will be inserted into the reservoir.
If the item is selected, then one of the existing items of the
reservoir is uniformly selected and replaced with the new item. The trick here is that,
if the probabilities of all items in the reservoir are already proportional to their
weights, then by selecting uniformly which item to replace, the probabilities of all items remain
proportional to their weight after the replacement.

\begin{figure}[ht]
\begin{center}
\begin{minipage}[b]{0.9\linewidth}
{\footnotesize
\underline{\bf Algorithm A-Chao (sketch)}\vspace{0.1cm} \\
{\bf Input : } Item $v_k$ for $m < k \leq n$ \\
{\bf Output : }  A WRS-N-P sample of size $m$ \\
\begin{tabular}{ll}
{\bf 1 : } & Calculate the probability $p_k = w_k/(\sum_{i=1}^k w_i)$ for item $v_k$ \\
{\bf 2 : } & Decide randomly if $v_k$ will be inserted into the reservoir \\
{\bf 3 : } & \hspace{0,5cm} if No, do nothing. Simply increase the total weight \\
{\bf 4 : } & \hspace{0,5cm} if Yes, choose uniformly a random item from the \\
        & \hspace{0,5cm} reservoir and replace it with $v_k$ \\
\end{tabular}
}
\caption[Algorithm A-Chao]{A sketch of Algorithm A-Chao.  We assume that all the positions of
the reservoir are already occupied and that all item weights are feasible.}
\label{fig:A-Chao}
\end{minipage}
\end{center}
\end{figure}
The main approach of A-Chao is simple, flexible and effective. There are however
some complications inherent to problem WRS-N-P that have to be addressed.
As shown in Example~\ref{exa:infeasible}, an instance of WRS-N-P may temporarily not be feasible,
in case of data streams, or may not be feasible at all. This happens when the (current) population contains
one or more overweight items, i.e., items each of which has a relative weight greater than $1/m$.
The main idea to handle this case, is to sample each overweight item with probability $1$. Thus, each
overweight item automatically occupies a position in the reservoir. The remaining positions are
assigned with the normal procedure to the feasible items. In case of sampling over a data
stream, an initially infeasible (overweight) item may later become feasible as more items arrive. Thus,
with each new item arrival the relative weights of the infeasible items are updated
and if an infeasible item becomes feasible it is treated as such.
Appropriate procedures to initialize the reservoir and to handle the overweight items 
are described in~\cite{Ch82}.

\subsection{A-ES}
\label{sec:A-ES}
The algorithm of Efraimidis and Spirakis~\cite{ES06}, which we call A-ES,
is a sampling scheme for problem WRS-N-W.
In A-ES, each item $v_i$ of the population V independently
generates a uniform random number $u_i \in (0,1)$ and calculates a key
$k_i = {u_i}^{1/w_i}$. The items that possess the $m$
largest keys form a weighted random sample.
We will use the reservoir-based version of A-ES, where the algorithm
maintains a reservoir of size $m$ with the items with $m$ largest keys.

The basic principle underlying algorithm A-ES is the
remark that a uniform random variable can be ``amplified''
as desired by raising it to an appropriate power (Remark~\ref{rem:exp-uniform}).
A high level description of algorithm A-ES is shown in Figure~\ref{fig:A-ES}.

\begin{remark}(\cite{ES06})
\label{rem:exp-uniform} Let $U_1$ and $U_2$ be independent random
variables with uniform distributions in [0,1]. If $X_1 =
(U_1)^{1/w_1}$ and $X_2 = (U_2)^{1/w_2}$, for $w_1, w_2 > 0$, then
$P[X_1 \leq X_2] = \displaystyle \frac{w_2}{w_1+w_2}$.
\end{remark}

\begin{figure}[ht]
\begin{center}
\begin{minipage}[b]{0.9\linewidth}
{\footnotesize
\underline{\bf{Algorithm A-ES} (High Level Description)}\vspace{0.1cm}\\
{\bf Input : } A population V of $n$ weighted items \\
{\bf Output : } A WRS-N-W sample of size $m$ \\
{\bf 1: } For each $v_i \in V$, $u_i = random(0,1)$ and $k_i =
u_i^\frac{1}{w_i}$ \\
{\bf 2: } Select the $m$ items with the largest keys $k_i$
}
\caption[Algorithm A-ES]{A high level description of Algorithm A-ES.}
\label{fig:A-ES}
\end{minipage}
\end{center}
\end{figure}

\subsection{Algorithm A-Chao with Jumps}
A common technique to improve certain reservoir-based sampling
algorithms is to change the random experiment used in the sampling procedure.
In normal reservoir-based sampling algorithms, a random experiment is
performed for each new item to decide
if it is inserted into the reservoir. In random sampling with jumps instead,
a single random experiment is used to directly decide which will be the next
item that will enter the reservoir.
Since each item that is processed will be inserted with some probability
into the reservoir, the number of items that will be skipped until the
next item is selected for the reservoir, is a random variable.
In uniform random sampling it is possible to generate an exponential
jump that identifies the next item of the population that will enter
the reservoir~\cite{De86}, while in~\cite{ES06} it is shown that
exponential jumps can be used for WRS with algorithm A-ES.

In this work, we show that a jumps approach can be used for algorithm A-Chao too, 
albeit in a slightly more complicated way than for algorithm A-ES. The reason is
that in WRS-N-W the probability that an item will be the next item
that will enter the reservoir can be directly obtained from its weight and
the total weight of the items preceding it, while in WRS-N-P the respective probabilities
have to be computed.

Assume for example a typical step of algorithm A-Chao. A new item $v_i$ has just arrived
and with probability $p_i$ it will be inserted into the reservoir.
The probability that $v_i$ will not be selected, but the next item, $v_{i+1}$, is selected,
is $(1-p_i) \, p_{i+1}$. In the same way the probability that items $v_i$ and
$v_{i+1}$ are not selected and that item $v_{i+2}$ is selected
is $(1-p_i) \, (1-p_{i+1}) \, p_{i+2}$.
Clearly, if the stream continues with an infinite number of items then
with probability $1$ some item will be the next item that will enter
the reservoir. Thus, we can generate a uniform random number $u_j$ in $[0,1]$
and add up the probability mass of each new item until the accumulated probability
exceeds the random number $u_j$. The selected item is then inserted into
the reservoir with the normal procedure of algorithm A-Chao.

The main advantage of using jumps in reservoir-based sampling algorithms
is that, in general, the number of random number generations can be dramatically
reduced. For example, if the item weights are independent random variables with
a common distribution, then the number of random numbers is reduced from $O(n)$
to $O(m \log (n/m))$, where $n$ is the size of the population~\cite{ES06}.
In contexts where the computational cost for qualitative random number generation is
high, the jumps versions offer an efficient alternative for the sampling
procedure. From a semantic point of view, the sampling procedures with and without 
jumps are identical.

\section{Algorithms for WRS Problems}
\label{sec:alg}
Both core algorithms, A-Chao and A-ES, are efficient and flexible and can be used to solve
fundamental but also more involved random sampling problems.
We start with basic WRS problems that are directly solved by A-Chao and A-ES.
Then, we present sampling schemes for two WRS problems with a bound on the number of
replacements and discuss the sampling problem in the presence of stream evolution.

\subsection{Basic problems}
\begin{itemize}
\renewcommand{\labelitemi}{$\circ$}
\item {\bf Problem WRS-N-P:} The problem can be solved with algorithm P-Chao.
In case no overweight items appear in the data stream, the cost to process each item
is $O(1)$ and the total cost for the whole population is $O(n)$.
The complexity of handling overweight items is higher. For example, if a heap
data structure is used to manage the current overweight items, then
each overweight item costs $O(\log m)$. An adversary
could generate a data stream where each item would be initially (at the time
it is fed to the sampling algorithm) overweight
and this would cause a total complexity of $\Theta(n \log m)$ to process
the complete population. However, this is a rather extreme example and in
reasonable cases the total complexity is expected to be linear on $n$.
\item {\bf Problem WRS-N-W:} The problem can be solved with algorithm A-ES.
The reservoir-based implementation of the algorithm requires $O(1)$ computational steps
for each item that is not selected and $O(\log m)$ for each item that
enters the reservoir (if, for example, the reservoir is organized as
a heap). In this case too, an adversary can prepare a sequence that
will require $O(n \log m)$ computational steps. In common cases,
the cost for the complete population will be $O(n) + O(m \log(n/m)) O(\log m)$,
which becomes $O(n)$ if $n$ is large enough with respect to $m$.
\item {\bf Problem WRS-R:} In WRS with replacement the population remains unaltered after
each item selection. Because of this, WRS-R-P and WRS-R-W coincide and we call the problem
simply WRS-R.
In the data stream version, the problem can be solved by running concurrently
$m$ independent instances of WRS-N-P or WRS-N-W, each with sample size $m'=1$.
Both algorithms A-Chao and A-ES in both their versions, with and without jumps,
can efficiently solve the problem. In most cases, the version with jumps
of A-Chao or A-ES should be the most efficient approach.

Note that sampling with replacement is not equivalent to running the experiment
on a population $V'$ with $m$ instances of each original item of $V$. The sample
space of the later experiment would be much larger than in the case with replacement.
\end{itemize}

\subsection{Sampling with a bounded number of replacements}
\begin{figure}[!htb]
\begin{minipage}[b]{\linewidth} % A minipage that covers half the page
\centering
% \fbox{
% \includegraphics[width=0.6\textwidth]{wrs-k.eps}
 \includegraphics[width=0.6\textwidth]{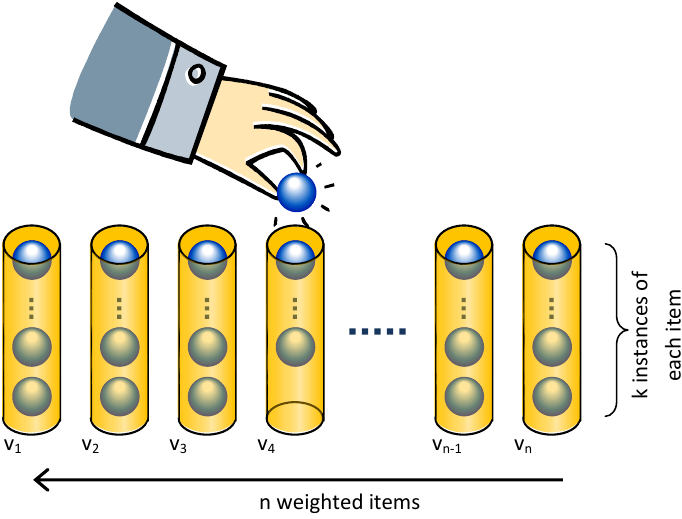}
% }
\caption{WRS-k-W, n weighted items with k instances of each item.} \label{fig:wrs-k}
\end{minipage}
\end{figure}

We consider weighted random sampling from populations where each item can be
replaced at most a bounded number of times (Figure~\ref{fig:wrs-k}).
An analogy would be to randomly select $m$ products from an automatic
selling machine with $n$ different products and $k$ instances of each product.
The challenge is of course that the weighted random sample has to be generated in
one-pass over an initially unknown population.

\begin{itemize}
\item {\bf Problem WRS-k-W:}
Sampling from a population of $n$ weighted items where each item can be selected up to $k \leq m$ times.
The weights of the items are used to determine the probability that each item is selected at each step. 

A general solution, in the sense that each item may have its own multiplicity $k_i \leq k$,
is to use a pipeline of $m$ instances of a A-ES, where each instance will generate a 
weighted random sample of size $1$. Note that either algorithm A-Chao or A-ES can be used, 
because for samples of size $1$ the outcomes of the two algorithms are equivalent. 
If the first instance is at item $\ell$, then each other instance is one item behind
the previous instance. Thus, an item of the population is first processed by instance $1$,
then by instance $2$, etc. If at some point the item has been selected $k_i$ times, then the
item is not processed by the remaining instances and the information up to which instance
the item has been processed is stored. If the item is replaced in a reservoir at a later step,
then it is submitted to the next instance of A-ES. Note that in this approach, some items
might be processed out of their original order. This is fine with algorithm A-ES
(both A-ES and A-Chao remain semantically unaffected by any ordering of the population)
but may be undesirable in certain applications.
\end{itemize}
\subsection{Sampling Problems in the Presence of Stream Evolution}

A case of reservoir-based sampling over data streams
where the more recent items are favored in the sampling process
is discussed in~\cite{Ag06}. While the items do not have weights and
are uniformly treated, a temporal bias function is used
to increase the probability of the more recent items to belong
to the random sample. Finally, in~\cite{Ag06}, a particular biased
reservoir-based sampling scheme is proposed and the problem of
efficient general biased random sampling over data streams is stated
as an open problem.

In this work, we have brought to the fore algorithms A-Chao and A-ES,
which can efficiently solve WRS over data streams where each
item can have an arbitrary weight.
This should provide an affirmative answer to the open problem posed in~\cite{Ag06}.
Moreover, the particular sampling procedure presented in~\cite{Ag06} is a
special case of algorithm A-Chao.

Since algorithms A-Chao and A-ES can support arbitrary item weights,
a bias favoring more recent items can be encoded into the weight
of the newly arrived item or in the weights of the items already in
the reservoir.
Furthermore, by using algorithms A-Chao and A-ES the sampling process
in the presence of stream evolution can also support weighted items.
This way the bias of each item may depend on the item weight and
how old the item is, or on any other factor that could be taken into account.
Thus, the sampling procedure and/or the corresponding applications in~\cite{Ag06}
can be generalized to items with arbitrary weights and other, temporal
or not, bias criteria.

The way to increase the selection probability of a newly arrived item
is very simple for both algorithms, A-Chao and A-ES.
\begin{itemize}
\item {\bf A-Chao:} By increasing the weight of the new item.
\item {\bf A-ES:} By increasing the weight of the new item or decreasing the
weights of the items already in the reservoir.
\end{itemize}

\section{An Abstract Data Structure for WRS}
\label{sec:impl}
\begin{figure}[!htb]
\begin{minipage}[b]{\linewidth} % A minipage that covers half the page
\centering
 \includegraphics[width=0.7\textwidth]{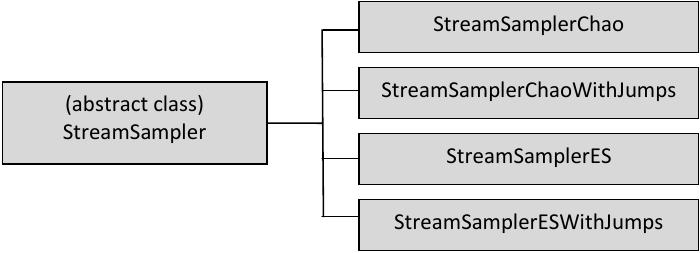}
\caption{The class hierarchy for sampling over data streams.} \label{fig:class}
\end{minipage}
\end{figure}

We designed an abstract class StreamSampler with the
methods feedItem() and getSample(), and a set of auxiliary classes for
the weighted items to capture the basic
functionality for weighted random sampling over data streams (Figure~\ref{fig:class}).
Then, we developed descendant classes
that implement the functionality of the StreamSampler class
for algorithms A-Chao and A-ES, both with and without jumps.
The descendant classes are StreamSamplerChao, StreamSamplerES,
StreamSamplerESWithJumps and StreamSamplerChaoWithJumps~\cite{WRS10}.

\begin{figure}[!ht]
\centering
\subfloat[Measurements for m=200 and n ranging from 5000 to 100000.]
{\label{fig:wrs2D}\includegraphics[width=0.48\textwidth, height=0.30\textwidth]{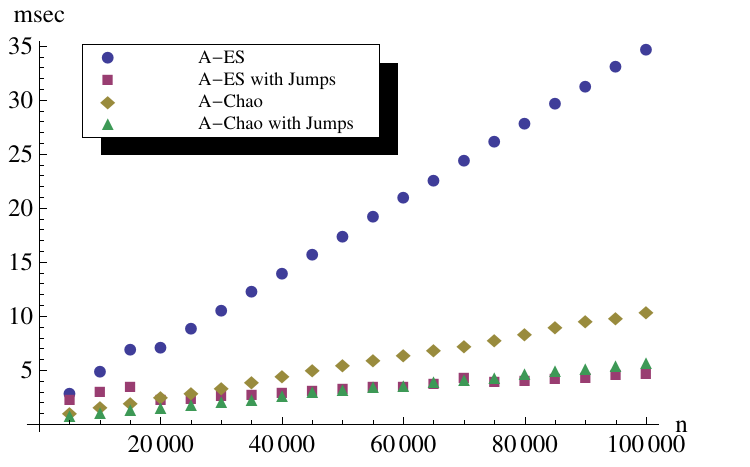}}
\hspace{0.02 \textwidth}
\subfloat[The complexity of A-ES for m ranging from 50 to 750 and n from 1000 to 6000.]
{\label{fig:wrsES3D}\includegraphics[width=0.48\textwidth, height=0.30\textwidth]{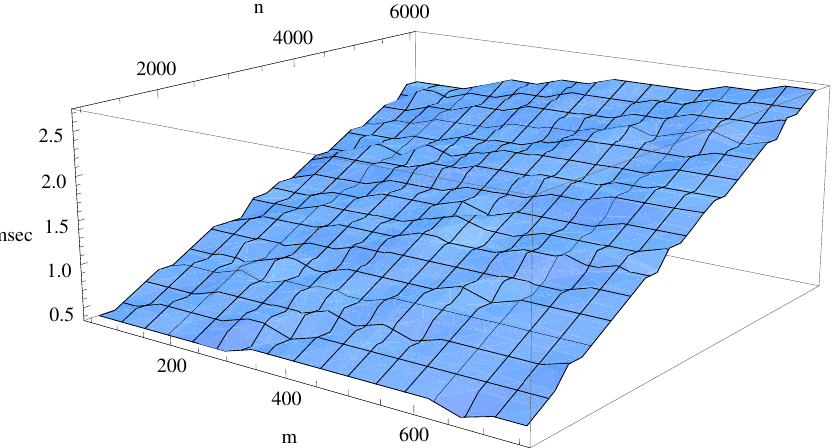}}
\caption{Time measurements of the WRS sampling algorithms.} \label{fig:meas}
\end{figure}

\noindent
Preliminary experiments with random populations (with uniform random item weights)
showed that all algorithms scale linear on the population size and at most linear
on the sample size. Indicative measurements are shown in Figure~\ref{fig:meas}.
While there is still room for optimization of the implementations of the algorithms,
the general behavior of the complexities is evident in the graphs. The experiments
have been performed on the Sun Java 1.6 platform running on an Intel Core 2 Quad CPU-based PC
and all measurements have been averaged over 100 (at least) executions.

\section{The Role of Weights}
\label{sec:weights}
The problem classes WRS-P and WRS-W differ in the way the item weights
are used in the sampling procedure.
In WRS-P the weights are used to directly determine the final selection probability
of each item and this probability is easy to calculate.
On the other hand, in WRS-W the item weights are used to determine the selection
probability of each item in each step of a supposed sequential sampling procedure.
In this case it is easy to study each step of the sampling
procedure, but the final selection probabilities of the items seem to be hard to
calculate. In the general case, a complex expression has to be
evaluated in order to calculate the exact inclusion probability of each item
and we are not aware of an efficient procedure to calculate this expression.
An interesting feature of random samples generated with WRS-W is that
they support the concept of order for the sampled items.
The item that is selected first or simply has the largest key (algorithm A-ES)
can be assumed to take the first position, the second largest the second position etc.
The concept of order can be useful in certain applications. We illustrate the two
sampling approaches in the following example.

\begin{example}{On-line advertisements.}
A search engine shows with the results of each query a set
of $k$ sponsored links that are related to the search query.
If there are $n$ sponsored links that are relevant to a query
then how should the set of $k$ links be selected? If all
sponsors have paid the same amount of money then any uniform
sampling algorithm without replacement can solve the problem.
If however, every sponsor has a different weight then how
should the $k$ items be selected?
Assuming that the $k$ positions are equivalent in ``impact'',
a sponsor who has the double weight with respect to another
sponsor may expect its advertisement to appear twice as often in the results.
Thus, a reasonable approach would be to use algorithm A-Chao
to generate a WRS-N-P of $k$ items.
If however, the advertisement slots are ordered based on their impact,
for example the first slot may have the largest impact, the second
the second largest etc., then algorithm A-ES may provide the
appropriate solution by generating a WRS-N-W of $k$ items.
\end{example}

When the size of the population becomes large with respect to the
size of the random sample, then the differences in the selection probabilities
of the items in WRS-P and WRS-W become less important. The reason is that if the
population is large then the change in the population because of the removed items
has a small impact and the sampling procedure converges to random sampling
without replacement. As noted earlier, in random sampling with replacement
the two sampling approaches coincide.

\section{Discussion}
\label{sec:disc}
We presented a comprehensive treatment of WRS over data streams
and showed that efficient sampling schemes exist for fundamental
but also more specialized WRS problems.
The two core algorithms, A-Chao and A-ES have been proved efficient
and flexible and can be used to build more complex sampling schemes.

%\vspace{0.5cm}
%\noindent
%{\bf Acknowledgements.}
%The present work was supported in part by the project ATLAS (Advanced Tourism Planning), GSRT/CO-OPERATION/11SYN-10-1730,
%and by national ETAA funds.

%\bibliography{algo2015R1}

\end{document}